\documentclass[aps,prd,preprint,superscriptaddress,showpacs,floatfix,nobibnotes]{revtex4-1}
\usepackage{float}
\usepackage{latexsym}
\usepackage{amsmath}
\usepackage{amssymb}
\usepackage{graphicx}
\usepackage{setspace}
\usepackage{longtable}
\usepackage{enumitem}
\usepackage[bottom]{footmisc}
\usepackage{slashed}
\usepackage{color}

\usepackage{bm}
\usepackage{epsfig}
\newcommand{\bea}{\begin{eqnarray}}
\newcommand{\eea}{\end{eqnarray}}

\newcommand{\nn}{\nonumber}

\begin{document}

\title{Extension of the validity of proper time expansions in the study of glasma dynamics}

\author{Margaret E. Carrington}
\affiliation{Department of Physics, Brandon University,
Brandon, Manitoba R7A 6A9, Canada}
\affiliation{Winnipeg Institute for Theoretical Physics, Winnipeg, Manitoba, Canada}

\author{Bryce T. Friesen}
\affiliation{Department of Physics, University of Toronto, Toronto, Ontario M5S 1A7, Canada}

\author{Doug Pickering}
\affiliation{Department of Mathematics, Brandon University,
Brandon, Manitoba R7A 6A9, Canada}

\author{Shane Sangster}
\affiliation{Department of Mathematics, Brandon University,
Brandon, Manitoba R7A 6A9, Canada}

\author{Kaene Soopramania}
\affiliation{Department of Physics, Brandon University,
Brandon, Manitoba R7A 6A9, Canada}

\date{April 07, 2026}

\begin{abstract}

The earliest phase of an ultrarelativistic heavy ion collision can be described as a highly populated system of gluons called glasma. The system's dynamics is governed by the classical Yang-Mills equation. Solutions can be found at early times using a proper time expansion. Since the expansion parameter is the proper time, this method is necessarily limited to the study of early time dynamics. In addition compute time and memory limitations restrict practical calculations to no more than eighth order in the expansion. The result is that the method produces reliable results only for very early times. In this paper we explore three different methods to increase the maximum time that can be reached. We use Pad\'e approximants to extrapolate analytic results calculated previously at eighth order in the proper time expansion \cite{Carrington:2023nty}.  We also use an analytic approximation based on the assumption that there are two widely separated ultraviolet scales \cite{Li:2016eqr} to extend the proper time expansion to 20th order. Lastly we develop a machine learning method to learn an expression for the transverse-longitudinal pressure anisotropy using symbolic regression. 
We find that, depending slightly on the quantity being calculated, the latest time for which reliable results are obtained can be extended approximately 1.5 times (from  $\sim0.05$~fm/$c$ using previous methods to about $0.08$~fm/$c$). 

\end{abstract}

\maketitle
\newpage

\section{Introduction}

In the Colour Glass Condensate (CGC) effective theory, the earliest phase of a heavy ion collision is a system of highly populated gluon fields called glasma (see, for example, the review \cite{Gelis:2010nm}). The lifetime of the glasma phase is only a fraction of a fm/$c$ but the development of a better understanding of glasma dynamics is important for several reasons. In previous work we have used a proper time expansion to obtain an analytic result for the CGC energy-momentum tensor (EMT) and shown that several phenomena observed through characteristics of final states of heavy ion collisions have their origin in the glasma phase \cite{Carrington:2023nty,Carrington:2020ssh,Carrington:2021qvi,Carrington:2024vpf,Carrington:2020sww,Carrington:2022bnv,Carrington:2021dvw,Carrington:2024utf,Carrington:2025xws}. The glasma is also important because it provides the initial conditions for the subsequent hydrodynamic evolution of the system.
In this paper we consider different methods to extend the range of proper time over which our calculations are valid. 

This paper is organized as follows. In section \ref{sec-method} we explain how we calculate the glasma EMT using a proper time expansion. Time and memory constraints limit these calculations to eighth order in the expansion. The radius of convergence of the expansion at eighth order is about $0.05-0.06$~fm/c, depending on the quantity that is being calculated. Since this time is so small, in particular much less than the time at which we expect the system to reach the hydrodynamic regime, it is of interest to develop methods to increase the radius of convergence of the expansion. We note that there are approaches to study the onset of hydrodynamics based on the finding that
non-equilibrium systems tend to evolve along a hydrodynamic attractor trajectory, see \cite{Heller:2015dha,Almaalol:2020rnu} and the review \cite{Jankowski:2023fdz}.
In section \ref{sec-ming} we show how to extend the order of the proper time expansion from eighth all the way to 20th order using an additional approximation. The method assumes that there are two distinct ultraviolet scales in the problem, the scale that sets the limit of the classical glasma approach, and the typical transverse momentum scale of the saturated gluons. The approximation assumes that these two scales are well separated and that one can therefore introduce a second expansion in their ratio. This assumption greatly reduces the total number of terms in the final expression for the EMT and makes it possible to extend calculations to higher orders. We will show that for some quantities this approximation works well but for others it does not capture the relevant physics. 
In section \ref{sec-pade} we explore the use of Pad\'e approximants to extend our full eighth order results to higher values of $\tau$. We show that using general approximants together with simple physically motivated contraints on the coefficients we obtain a smooth extrapolation of the energy and pressure anisotropy to much larger times. These results are largely independent of the form of the approximant.
In section \ref{sec-ML} we study the use of machine learning techniques to determine coefficients at higher orders in the proper time expansion. As a test of the method we show that for the transverse-longitudinal pressure anisotropy we can learn a function that is a polynomial and has the known eighth order coefficient, from data constructed from the sixth order results. We then use the same method learn the next two higher order coefficients and give an estimate of the errors in these results. 

We use  natural  units with $c = \hbar = k_B =1$. Our conventions for indices are: Greek letters $\mu, \nu, \rho, \dots$ for components of four-vectors, Latin letters $i,j,k, \dots$ for components of two-component vectors transverse to the collision axis, and Latin letters from the beginning of the alphabet $a,b, c \dots$ for colour components of the SU($N_c$) gauge group. We use three coordinate systems: Minkowski $(t, z, x, y)$, light-cone $(x^+,x^-, x, y)$ and Milne $(\tau, \eta, x, y)$, where $x^\pm = (t\pm z)/\sqrt{2}$, $\tau=\sqrt{t^2-z^2}$ and $\eta=\ln(x^+/x^-)/2$. 

\section{Summary of computational method}
\label{sec-method}

We calculate the EMT in terms of chromodynamic electric and magnetic fields averaged over the colour configurations of the colliding nuclei. We use a method called a proper time expansion \cite{Fries:2005yc,Fukushima:2007yk,Fujii:2008km,Chen:2015wia,Fries:2017ina} that is designed to describe the earliest phase of relativistic heavy-ion collisions. The procedure we use is described in detail in \cite{Carrington:2020ssh,Carrington:2021qvi,Carrington:2024vpf}. 
The method is based on an expansion of the chromodynamic potentials in powers of the proper time which is treated as a small parameter.  This expansion allows one to solve the classical Yang-Mills equations iteratively. We obtain an analytic result for the EMT which is valid only at very early times. 
We summarize the method below, more details can be found in \cite{Carrington:2023nty,Carrington:2021qvi,Carrington:2020ssh}. 

We consider a collision of two heavy ions moving with the speed of light towards each other along the $z$ axis and colliding at $t=z=0$. The vector potential of the gluon field is described with the ansatz \cite{Kovner:1995ts} 
\bea
A^+(x) &=& \Theta(x^+)\Theta(x^-) x^+ \alpha(\tau,\vec x_\perp) \label{ansatz}
\\
A^-(x) &=& -\Theta(x^+)\Theta(x^-) x^- \alpha(\tau,\vec x_\perp) 
\nn \\ \
A^i(x) &=& \Theta(x^+)\Theta(x^-) \alpha_\perp^i(\tau,\vec x_\perp)
+\Theta(-x^+)\Theta(x^-) \beta_1^i(x^-,\vec x_\perp)
+\Theta(x^+)\Theta(-x^-) \beta_2^i(x^+,\vec x_\perp) \nn
\eea
where the functions $\beta_1^i(x^-,\vec x_\perp)$ and $\beta_2^i(x^+,\vec x_\perp)$ represent the pre-collision potentials and $\alpha(\tau,\vec x_\perp)$ and $\alpha_\perp^i(\tau,\vec x_\perp)$ are the post-collision potentials. 
In the forward light-cone the vector potential satisfies the sourceless Yang-Mills equation. 
The sources enter through boundary conditions that connect the pre-collision and post-collision potentials. The boundary conditions are
\bea
\label{cond1}
&& \alpha^{i}_\perp(0,\vec{x}_\perp) = \alpha^{i(0)}_\perp(\vec{x}_\perp) 
= \lim_{\text{w}\to 0}\left(\beta^i_1 (x^-,\vec{x}_\perp) + \beta^i_2(x^+,\vec{x}_\perp)\right) \nn
\\
\label{cond2}
&& \alpha(0,\vec{x}_\perp) = \alpha^{(0)}(\vec{x}_\perp) 
= -\frac{ig}{2}\lim_{\text{w}\to 0}\;[\beta^i_1 (x^-,\vec{x}_\perp),\beta^i_2 (x^+,\vec{x}_\perp)] 
\eea
where the notation $\lim_{\text{w}\to 0}$ indicates that the width of the sources across the light-cone is taken to zero as the colliding nuclei are infinitely contracted. 

We find solutions valid for early post-collision times by expanding the Yang-Mills equations in the proper time $\tau$. These solutions are used to write the post-collision field-strength tensor and the EMT in terms of the initial potentials $\alpha(0, \vec x_\perp)$ and $\vec\alpha_\perp(0, \vec x_\perp)$ and their derivatives, which are then written in terms of the pre-collision potentials $\vec \beta_1(x^-,\vec x_\perp)$ and $\vec \beta_2(x^+,\vec x_\perp)$ and their derivatives using the boundary conditions (\ref{cond2}). 
We then use the Yang-Mills equations in the pre-collision region to write the pre-collision potentials in terms of the colour charge distributions of the incoming ions. The correlator of two pre-collision potentials from different ions is set to zero. 
All the physical quantities we study are constructed from the correlators for two potentials from the same ion
\bea
\delta^{ab} B_n^{ij}(\vec{x}_\perp,\vec y_\perp) \equiv 
\lim_{{\rm w} \to 0}  \langle \beta_{n\,a}^i(x^-,\vec x_\perp) \beta_{n\,b}^j(y^-,\vec y_\perp)\rangle  
\label{beta2B}
\eea
and their derivatives. The index $n\in(1,2)$ indicates one of the two incoming ions. 
The correlator is proportional to the surface density of colour charge of each ion which we denote $\mu_n(\vec x_\perp)$. We use a two-dimensional projection of a Woods-Saxon distribution onto the plane transverse to the collision axis
\bea
\label{def-mu2-2}
\mu(\vec x_\perp)  =  
\Big(\frac{A}{207}\Big)^{1/3}\frac{\bar\mu}{2a\ln(1+e^{R_A/a})} 
\int^\infty_{-\infty}\frac{dz}{1 + \exp\big[(\sqrt{\vec x_\perp^2 + z^2} - R_A)/a\big]} \,.
\eea
The parameters $R_A$ and $a$ give the radius and skin thickness of a nucleus of mass number $A$ and $\bar\mu$ is the value of $\mu(\vec x_\perp)$ at the center of the nucleus when $A=207$. We use $r_0=1.25$ fm and $a=0.5$ fm so that the radius of a nucleus with $A=207$ is $R_A=r_0 A^{1/3}= 7.4$ fm. 
To obtain analytic results for the correlators in eq.~(\ref{beta2B}) we perform a gradient expansion. We define the vector $\vec {\cal R} = (\vec x_\perp + \vec y_\perp)/2$ and expand the colour charge densities (\ref{def-mu2-2}) around the centers of colliding nuclei and we take into account only the first two terms in the expansion. To consider collisions with nonzero impact parameter we displace the centers of the two ions by $\pm \vec b/2$ where the impact parameter vector $\vec b$ is directed along the $x$-axis. We use $\mu_1(\vec R) \equiv \mu(\vec R-\vec b/2)$ and $\mu_2(\vec R) \equiv \mu(\vec R + \vec b/2)$ so that the position $\vec R=0$ is the center of the interaction region.

A detailed explanation of the method we use to calculate the correlator (\ref{beta2B}) and its derivatives can be found in ref.~\cite{Carrington:2020ssh}. 
The result depends on three scales. 

One is an ultraviolet regulator, denoted $\Lambda$, that sets the upper bound for the validity of the classical picture that underlies the glasma approach we are using. The dimensionless parameter $\Lambda\tau$ characterises the convergence of the proper time expansion. 

The surface charge density functions introduce another hard scale, $\bar\mu$ in eq.~(\ref{def-mu2-2}), which is conventionally called the McLerran-Venugopalan (MV) scale. The MV scale is related to the saturation scale, denoted $Q_s$, and we make the standard choice $\bar\mu = Q_s/g^4$ (some motivation can be found in \cite{Carrington:2020ssh}). We also define a dimensionless charge density function
\bea
\hat \mu_n(R_x,R_y) \equiv \frac{g^4}{Q_s^2}\mu_n(R_x,R_y) \,.
\label{hat-def0}
\eea

The third scale is an infrared regulator, denoted $m$, that is related to the confinement scale. 
The dimensionless parameters that determine the convergence of the gradient expansion are ratios of derivatives of the charge density and this scale. We define  dimensionless hatted derivatives as
\bea
\hat \mu_n^{(i,j)}(R_x,R_y) \equiv \frac{g^4}{Q_s^2} \frac{1}{m^{i+j}}   \frac{d^i}{dR_x^i}\frac{d^j}{dR_y^j}\mu_n(R_x,R_y) \,.
\label{hat-def}
\eea
The gradient expansion is valid in the region of the transverse plane where $\delta$ is less than one with the definition
\bea
\delta \equiv \frac{\hat \mu_n^{(i,j)}(R_x,R_y)}{\hat\mu_n(R_x,R_y) }\,.
\eea

The infrared regulator is always set to $m=0.2$ GeV.  We consider several different values of the two hard scales. All calculations are done with $A=207$ and at mid-rapidity ($\eta=z=0$). All distances are given in femtometer (fm).

\section{Method of Li \& Kapusta}
\label{sec-ming}

\subsection{Theory}

In the first part of this study we implement an additional approximation based on an idea of Li and Kapusta \cite{Li:2016eqr}.  
These authors proposed a method to exploit the fact that the two ultraviolet energy scales in the problem could be widely separated. As explained in the previous section, these scales are
$\Lambda$, which gives the upper bound for the transverse momenta of soft partons that can be effectively described as classical fields, and the saturation scale, $Q_s$, which describes the typical
transverse momentum scale of saturated gluons. 
In the CGC approach that we use the color sources are assumed
to be uncorrelated on the transverse plane, which requires that the resolution scale must be at least as small as $1/Q_s$, or that $\Lambda$ must be at least as large as $Q_s$. In our earlier work we made the conventional choice and set $\Lambda=Q_s$. 
The approximation of Li and Kapusta (LK) is to assume $\Lambda \gg Q_s \gg m$ (where $m$ is the confinement scale, as discussed in the previous section). 
We emphasize that the key to implementing the LK approximation is not to extract the dominant terms from the full calculation but rather to develop a method to identify which terms are dominant before doing the full calculation. We explain how to do this below. 

The result for the basic correlator in eq.~(\ref{beta2B}) is (see ref.~\cite{Carrington:2020ssh} for details)
\bea
&& \lim_{\vec x_\perp \to \;\vec y_\perp}B_n^{ij}(\vec x_\perp,\vec y_\perp) 
=\delta^{ij}g^2\frac{\mu_n(\vec R)}{8\pi}\left[\ln\left(\frac{\Lambda^2}{m^2}+1\right)-\frac{\Lambda^2}{\Lambda^2+m^2}\right] \nn\\
&& + \frac{g^2}{16\pi(\Lambda^2+m^2)} \left[\delta^{ij} \nabla^2\mu_n(\vec R)\frac{\Lambda^4}{6m^2(\Lambda^2+m^2)}\left(1+\frac{2m^2}{\Lambda^2+m^2}\right)
+ \nabla^i\nabla^j \mu_n(\vec R)\frac{\Lambda^2}{m^2}\right]\,.
\label{g-corr-res-basic}
\eea
Correlation functions involving derivatives of pre-collision potentials can be calculated with the same method that was used to obtain eq.~(\ref{g-corr-res-basic}) but they cannot be obtained directly from (\ref{g-corr-res-basic}) since the derivatives must be taken before the limit $\vec x_\perp \to \vec y_\perp$ is taken. There are two important features of all correlation functions that can be seen in (\ref{g-corr-res-basic}). The first is that the scale $Q_s$ enters through the charge density functions, which means that every correlator is proportional to $Q_s^2$. The second is that the parameter $m$ can only appear in two ways. The gradient expansion gives derivatives of the charge density function  accompanied by factors of $m$ that are absorbed into the hatted derivatives in eq.~(\ref{hat-def}), which have the same dimension as the original charge densities. All remaining factors of $m$ appear in the form of factors $(\Lambda^2+m^2)$ in denominators, or in arguments of logarithms.
In combination we have that if we use the LK assumptions $m/Q_s \ll 1$ and $Q_s/\Lambda \ll 1$ then the EMT can be expanded in the form
\bea
T^{\mu\nu} = Q_s^4 \sum_{n=0}  \; \tilde\tau^n \sum_{i=0} \left(\frac{Q_s^2}{\Lambda^2}\right)^i\sum_{j=0}c_{nij}  \left(\frac{m^2}{\Lambda^2}\right)^j \label{ming-exp}
\eea
where we have defined $\tilde\tau = \tau/\Lambda$. 
The coefficients $c_{nij}$ of the terms in this expansion  depend only on the dimensionless hatted charge densities in eqs.~(\ref{hat-def0}, \ref{hat-def}) and factors $\log(m/\Lambda)$. 

Since the LK assumption is $m/Q_s \sim Q_s/\Lambda \sim \epsilon$, we have that the leading order contributions in (\ref{ming-exp}) are proportional to $Q_s^4$. The next-to-leading order contributions are from terms $(i,j) = (1,0)$ and next-to-next-to-leading order comes from $(i,j) = (2,0)$ and $(0,1)$. The important point is that when we consider corrections of   order $\epsilon^{2k}$ to the leading order result, we are guaranteed that no contributions will be missed if we take the upper limit of the $i$-summation equal to $k$. This happens because enhancement factors of the form $\Lambda/m$ do not occur, due to the form of the correlators, as seen in eq.~(\ref{g-corr-res-basic}). 
Since all factors $Q_s$ come from the charge density functions, a term $\sim Q_s^{2k}$ must come from a product of $k$ correlators, which means that in the original expression there was a product of $2k$ pre-collision potentials. 
The conclusion is that at leading order there are no contributions to the EMT from products of more than four pre-collision potentials, and at next-to-leading order there are no contributions from terms with more than six pre-collision potentials. This results in an enormous simplification because applying Wick's theorem to products of large numbers of potentials involves huge combinatoric factors. For example, the number of possible contractions of 4 factors of pre-collision potentials is 3, but for 12 pre-collison potentials there are 10,395 possible contractions. 

We also note that if there is only one ultraviolet scale then since the EMT is given by purely classical field configurations the coupling dependence can be entirely factored out. When two ultraviolet scales are present some higher order effects can be taken into account with a running coupling constant. 
The coupling constant in the Yang-Mills equation which enters the correlator (\ref{beta2B}) is replaced by the running coupling $g(\Lambda)$ and the MV scale becomes $\bar\mu = Q_s/g^4 \to Q_s^2/g(Q_s)^4$. The running couplings are implemented using 
\bea
&& g(X) = \sqrt{4\pi\alpha(X)} \text{~~with~~}  \alpha(X) = 1/(2\beta\log(X/m)) \label{runn}
\eea
where $\beta=(11N_c-2N_f)/(12\pi)$. 

\subsection{Results}

We study 
the transverse-longitudinal pressure anisotropy by looking at the quantity 
\bea
A_{TL} = \frac{3(p_T-p_L)}{2p_T+p_L}\,.
\label{Atl-def}
\eea
The transverse and longitudinal pressures are initially far from each other because the system is produced in a highly unequilibrated state. We expect that as $\tau$ increases the system will start to isotropize. This means that the transverse and longitudinal pressures will approach each other and the measure $A_{TL}$ will decrease.  

\begin{figure}[h]
\begin{center}
\includegraphics[width=15cm]{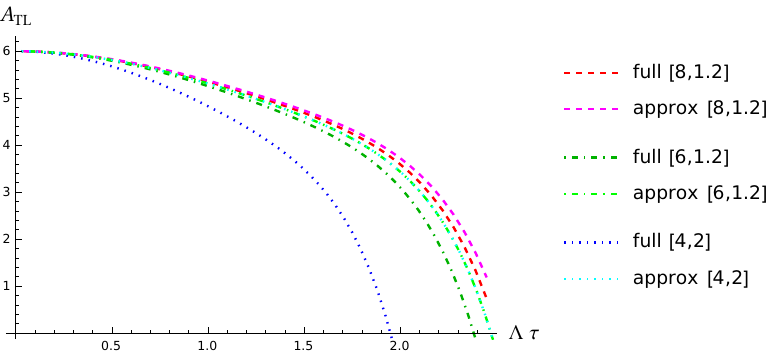}
\end{center}
\caption{The quantity $A_{TL}$ versus $\Lambda\tau$ in the full theory and the leading order approximation to it, at sixth order in the proper time expansion, with $R=b=0$, using different values of $(\Lambda,Q_s)$ (indicated by the numbers in the square brackets, in GeV). The light green line is partially hidden by the cyan line because they are so close together.
\label{ming-versus-full}}
\end{figure}

\begin{figure}[h]
\begin{center}
\includegraphics[width=11cm]{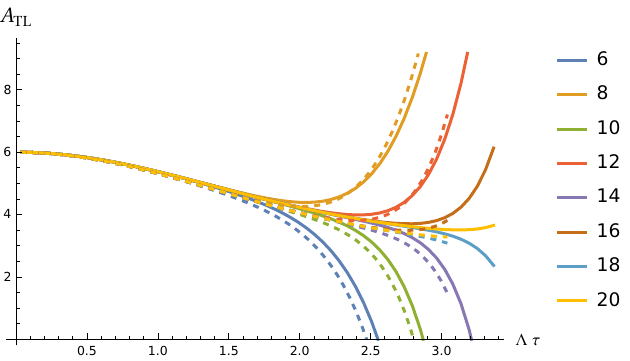}
\end{center}
\caption{The quantity $A_{TL}$ versus $\Lambda\tau$ using the leading order LK approximation at different orders in the proper time expansion using $Q_s=1.2$~GeV and two values of $\Lambda$: 8~GeV (solid) and 6~GeV (dotted) with $R=b=0$. 
\label{ming-to-20}}
\end{figure}

In fig.~\ref{ming-versus-full} we show $A_{TL}$ versus the proper time, in the full theory and the leading order LK approximation, for different values of $\Lambda$ and $Q_s$. The calculation is done at $R=0$ and for zero impact parameter, at sixth order in the proper time expansion. The figure shows that the LK approximation gives a good representation of the full theory when the two ultra-violet scales are well separated. 
In fig.~\ref{ming-to-20} we show the convergence of the proper time expansion up to 20th order, at leading order in the LK approximation. The figure shows that the radius of convergence of the expansion is significantly extended using the LK approximation.
From the graph we can say that results are reliable to $\Lambda\tau\approx 3.2$ with $\Lambda=8$ which gives an upper bound $\tau\approx 0.08$~fm. For comparison the full calculation at eighth order which can be trusted to only about $\tau Q_s\approx 0.5$ or $\tau\approx 0.05$~fm (see fig.~\ref{fig-ML2}). 

\begin{figure}[h]
\begin{center}
\includegraphics[width=11cm]{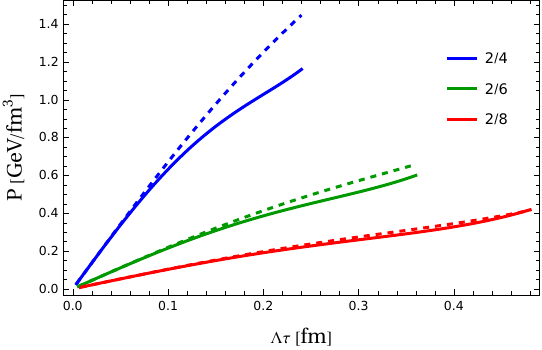}
\end{center}
\caption{The quantity $P$ versus $\Lambda\tau$ for different values of $Q_s/\Lambda$ as shown in the legend [in GeV]. The solid lines are the full calculation and the dotted lines are the approximation. 
\label{plot-P-ming}}
\end{figure}

We also calculate the radial flow which is characterized by the radial projection of the transverse Poynting vector $P=\hat R^i T^{i0}$ where $\hat R^i = R^i/|\vec R|$. The results are shown in fig.~\ref{plot-P-ming}. 
The calculation is done at fifth order in the proper time expansion, at $R=3$~fm and $\phi=0$ (in the reaction plane) and with $b=6$~fm. For radial flow the approximation of Li and Kapusta agrees fairly well with the full calculation for a wide range of values of $(Q_s,\Lambda)$. The problem is that results are fairly sensitive to the ratio of these values and become largely suppressed when $Q_s/\Lambda$ is less than one by any significant amount. This happens because the radial flow depends on the derivative terms of the source density functions, and these terms are suppressed by the approximation. The result is that the LK~approximation largely kills the effects of nuclear structure that we want to observe. 
In table~\ref{table-P} we show the value of $P$ at $t=0.06$~fm from the full calculation and from the approximation, and the percent by which their ratio is different from one. These results show clearly that when $P$ is large the approximation is not accurate, and when it is accurate $P$ is small. 
\begin{table}[H]
\begin{center}
\begin{tabular}{|c|c|c|c|c|}
\hline
~~ $Q_s$ ~~& ~~ $\Lambda$ ~~ & ~~  $P(0.06\text{~fm})$ ~~ & ~~ $P_{\rm approx}(0.06\text{~fm})$ ~~ &~~ $(P_{\rm approx}/P-1)\times 100\%$ ~~ \\
\hline\hline
  2. & 2. & 8.49 & 12.2 & 44.2 \\
  \hline
 2. & 4. & 2.16 & 2.63 & 22.1 \\
 \hline
 2. & 6. & 1.1 & 1.2 & 8.86\\
 \hline
 2. & 8. & 0.759 & 0.757 & -0.259 \\
 \hline
 \end{tabular}
 \end{center}
 \caption{The radial flow in GeV/fm$^3$ for different values of $Q_s$ and $\Lambda$ in GeV at $\tau=0.06$~fm. \label{table-P}}
\end{table}

\begin{figure}[H]
\begin{center}
\includegraphics[width=9cm]{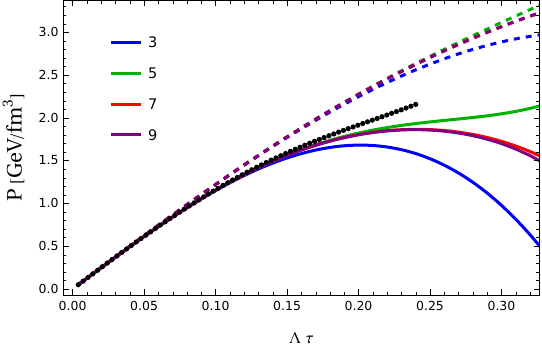}
\end{center}
\caption{The quantity $P$ versus $\Lambda\tau$ using the LK approximation at LO (dashed lines)and NLO (solid lines) with $\Lambda=4$~GeV and $Q_s=2.0$~GeV up to ninth order in the proper time expansion. The black line shows the result from the full EMT at fifth order. 
\label{ming-nlo}}
\end{figure}

An obvious way to extend the validity of the LK approximation is to include next-to-leading order (NLO) terms in the expansion in $\epsilon\sim Q_s/\Lambda$. The calculational advantage is greatly reduced, but one can test the importance of the NLO contributions.  In fig~\ref{ming-nlo} we show $P$ with $\Lambda=4$~GeV and $Q_s=2.0$~GeV at $R=3$~fm and $\phi=0$ and for $b=6$~fm. 
The black line shows the result from the full EMT at fifth order. The coloured lines show the LK approximation at LO (dashed lines) and NLO (solid lines) at orders 3,5,7 and 9 in the proper time expansion. The figure shows that the next-to-leading order approximation matches better with the full calculation, but the agreement is not greatly improved with $Q_s/\Lambda = 0.5$. 
Smaller values of $Q_s/\Lambda$ would obviously improve the accuracy of the approximation but, as discussed above, the physical effect we want to measure is not accurately described by the LK approximation when the parameter $\epsilon$ is taken too small. 

We conclude that the approximation of Li and Kapusta is not well suited to calculate quantities that depend on the derivatives of the source density functions, which means it cannot be used to study the effects of nuclear structure. 

\section{Pad\'e approximants}
\label{sec-pade}

In this section we study the use of Pad\'e approximants to extend the region of validity of the proper time expansion. We will consider the energy density and the pressure anisotropy as defined in equation (\ref{Atl-def}). Throughout this section we use $Q_s=2.0$~GeV. 

\subsection{Theory}
We construct Pad\'e approximants to data obtained from results calculated at eighth order in the proper time expansion. 
We consider only $R=b=0$. 
We use two different functions of the form
\bea
&& f_2(x) = \frac{a_0 + x^2 a_2}{b_0 + x^2 b_2} \nn \\
&& f_4(x) = \frac{(a_0 + x^2 a_2)(b_0 + x^2 b_2)}{(c_0 + x^2 c_2)(d_0+x^2 d_2)}\,.\nn 
\label{pade-fcns}
\eea
In all cases the coefficients are found using a least squares fit subject to the constraint  that the denominator does not have zeros. For the function $f_2(x)$ the constraint takes the form  $b_0b_2 > 0$ and for $f_4(x)$ we require $c_0c_2 >0$ and $d_0d_2 > 0$.

\begin{figure}[h]
\begin{center}
\includegraphics[width=12cm]{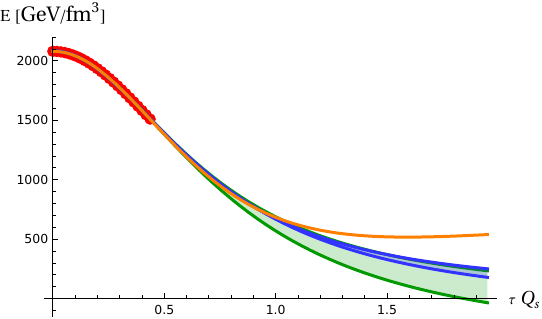}
\end{center}
\caption{The energy density at eight order (red markers) and the result from the second and fourth order Pad\'e approximants. The blue band shows the difference between the two approximants with maximum value of the data set to $\tau Q_s= 0.60$ and the green band is obtained with a maximum of 0.45. The orange curve is the fourth order approximant with coefficients determined by matching at the origin. 
\label{plot-pade}}
\end{figure}

For the energy density we select the data to fit using two different values of the maximum time. For the smaller value the reliability of the eighth order result is more certain, and for the larger value the range of data points that is used is bigger. 
The results are shown in fig.~\ref{plot-pade}. The green and blue bands shows the range of  values from the two different approximants when the upper limit of the time variable for the data to be fit is $\tau Q_s=0.45$ and $0.60$, respectively. In both cases the two approximants produce very similar results, and the blue band is completely contained within the green one. This demonstrates the stability of the method. The largest time at which the result can be trusted is $Q_s \tau\approx 1.5$ or $\tau\approx 0.15$~fm. For comparison the radius of convergence of the full calculation is about $\tau\approx 0.06$~fm (see fig.~1 in ref.~\cite{Carrington:2024vpf}). We also show (in orange) the result obtained from a fourth order approximant with coeffients obtained by matching to the Taylor expansion at the origin. This  curve matches the data well at early times but rises at later times which is unphysical. Since the fit is obtained using only information from the original function at $\tau\approx 0$ it is not expected to be as reliable at later times. 

\begin{figure}[h]
\begin{center}
\includegraphics[width=12cm]{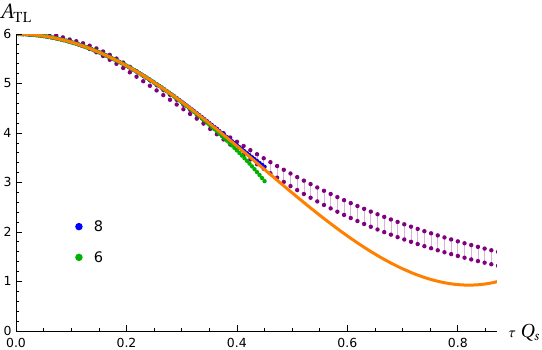}
\end{center}
\caption{The quantity $A_{TL}$ at sixth and eight order and the results from the second and fourth order Pad\'e approximants. The maximum difference between the four Pad\'es is indicated by the purple band.  The independent variable is $\tau Q_s$ with $Q_s=2.0$~GeV. The orange curve is the fourth order approximant with coefficients determined by matching at the origin.
\label{plot-padeA}}
\end{figure}

The Pad\'e of the quantity $A_{TL}$ is more sensitive to the choice of the upper bound on the time. The reason is that the proper time expansions at sixth and eighth order diverge in opposite directions (towards negative infinity at sixth order and positive infinity at eighth order). We therefore use 100 data points in the range $0< \tau Q_s < 0.45$, which is where the values and curvature of the sixth and eighth order data agree well. 
The results for the pressure anisotropy are shown in fig.~\ref{plot-padeA}. 
We have used the second and fourth order approximants with the sixth and eighth order data. The purple band indicates the maximum spread in the values of $A_{TL}$ from all four calculations. From the graph the result can be trusted to $Q_s\tau\approx 0.8$ or $\tau\approx 0.08$~fm which is similar to what we obtained from the LK approximation in the previous section. For comparison we show (in orange) the fourth order approximant with coefficients obtained by matching at the origin. As expected, the result is reliable at small times but displays unphysical behaviour (an increase in the pressure anisotropy) at later times.  

\section{Machine learning}
\label{sec-ML}

In this section we explore the use of machine learning techniques to extend the range of validity of the proper time expansion. We consider only the quantity $A_{TL}$ at the center of the collision where the derivatives of the charge density functions are all zero. We take $\Lambda=Q_s=2.0$~GeV and $m=0.2$~GeV and we use  the dimensionless variable $\tilde\tau = Q_s\tau$. 

The proper time expanded expression for $A_{TL}$ has the generic form
\bea
A_{TL} = \frac{n_0+n_2\tilde\tau^2+n_4\tilde\tau^4+n_6\tilde\tau^6+n_8\tilde\tau^8 +\dots}{d_0+d_2\tilde\tau^2+d_4\tilde\tau^4+d_6\tilde\tau^6+d_8\tilde\tau^8 +\dots}\,
\label{A-ana}
\eea
where the $n_i$ and $d_i$ are real numbers. 
We will use that the ratios of the coefficients are 
\bea
\frac{n_i}{d_i} = \frac{9i}{2}+6\,.
\label{magic}
\eea
We have not been able to derive this expression but we have checked that it is satisfied analytically to eighth order. 

The numerical results for the coefficients $n_i$ at eighth order are 
\bea
\{n_0,n_2,n_4,n_6,n_8\} = \{95.8636,  -445.139,  1024.37,  -1625.86,  2119.93\}\,.
\label{us8}
\eea
We use the package \textbf{PySR }\cite{Cranmer}  to learn an improved function that will be valid at later times. The allowed operations are addition and multiplication. Powers are not explicitly included but they are generated by these allowed operations. The code is given a  two dimensional array of 1200 points of the form $(\tilde\tau_{\rm value},y)$ where $y$ is the computed numerator. We use 1200 values of $\tilde\tau$  $\in(0,0.5)$ which is the range where the eighth order results are reliable.
To obtain the $y$ values we use the function $n_0+n_2\tilde\tau^2+n_4\tilde\tau^4+r_6 n_6\tilde\tau^6+r_8 n_8\tilde\tau^8+r_{10} n_{10}\tilde\tau^{10}+ r_{12} n_{12}\tilde\tau^{12}$. The coefficients $\{n_0,n_2,n_4,n_6,n_8\}$ are taken from eq.~(\ref{us8}), the numbers $\{r_6,r_8,r_{10},r_{12}\}$ are random numbers taken from a normal distribution with standard deviation $0.1$, and we use $n_{10}=-2400\pm 120$ and $n_{12}=2600\pm 130$ which are educated guesses obtained by looking at the lower order coefficients in (\ref{us8}). The code is given this array and asked to make 1000 attempts to learn the numerator of $A_{TL}$ in eq.~(\ref{A-ana}) by repeatedly modifying its trial function and evaluating the success of each modification using a least squares fit. An example of the output is
\bea
34.4 - ((x  (x  (x + 0.0427)  0.833 + 0.546)  (-29.8 x + 
          -5.13 x + 10.9) - 7.59)  (x - 0.66)  88.8 x - 
    61.5) \nonumber
\eea where $x=\tau^2$, which gives a polynomial of order $\tau^{12}$ when expanded. This process is repeated 500 times and many of the expressions produced are very long. 
Occasionally the machine produces anomalous results which are eliminated by removing results for which the absolute value of any coefficient is greater than $10^{7}$.
We calculate the median of the remaining values and estimate the error from the median deviation. 
We emphasize that since the initial data set in each of the 500 trials is constructed using $4\times 1200$ independent random numbers, it is extremely unlikely that the convergence of the results is a numerical artifact. Similarly the possible effect of input bias is largely eliminated using our method to construct the initial data. 

As a test of the method we learn a function using only the known values of the first four coefficients. The result is a polynomial with the coefficient of the eighth order term $[n_8]_{\text ML} = 2120.0 \pm 5$which agrees extremely well with the value 2119 in eq.~(\ref{us8}). When we use all of the known coefficients, the machine learned results for the 10th and 12th order coefficients are
\bea
&& [n_{10}]_{\rm ML} =-2394\pm 47 \nn \\
&& [n_{12}]_{\rm ML} = 2606\pm 76 \,.
\label{coes-10-12}
\eea
The uncertainties of the 10th and 12th order coefficients are approximately 1.9\% and 2.9\%, respectively. 

\begin{figure}[h]
\begin{center}
\includegraphics[width=7cm]{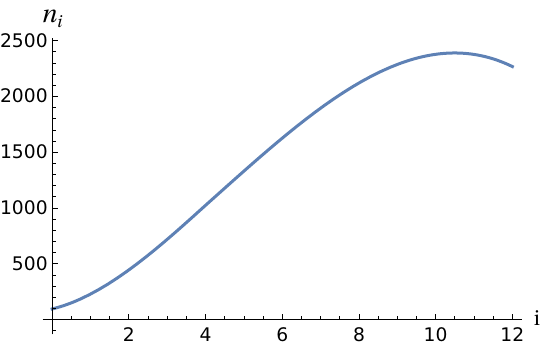}
\end{center}
\caption{The absolute value of the coefficients $n_i$ extrapolated from 8th to 12th order. 
\label{fig-extrapolate}}
\end{figure}

\begin{figure}[h]
\begin{center}
\includegraphics[width=8.1cm]{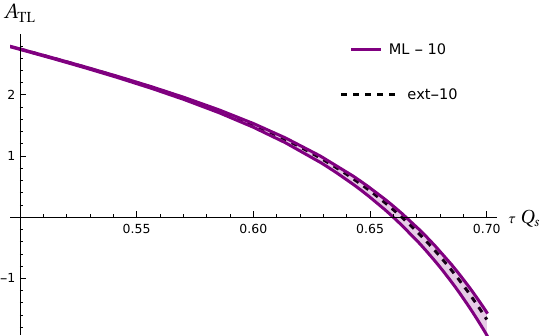}
\includegraphics[width=8.1cm]{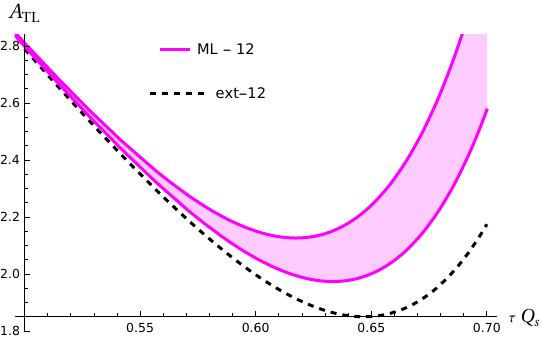}
\end{center}
\caption{$A_{TL}$ versus $\tau Q_s$ at 10th order (left) and 12th order (right). The black dotted line is the extrapolated result and the coloured bands show the results obtained with machine learning. 
\label{fig-ML1}}
\end{figure}

For comparison we calculate these coefficients by extrapolating the absolute values of the coefficients $n_i$. To do this we make an 2-dimensional array $\{i,|n_i|\}$ with the eight known results for the numerator coefficients and use cubic splines to construct an interpolating function (we have checked that the results are almost completely independent of the interpolation method). This gives $[n_{10}]_{\rm ext}=-2376.88$ and $[n_{12}]_{\rm ext}=2267.03$. The extrapolated value of $n_{10}$ lies within the error bar of the learned coefficient, but the result for $n_{12}$ is much too low. This happens because the extrapolated curve flattens and eventually dips down as $i$ increases, as shown in fig.~\ref{fig-extrapolate}. The failure is not unexpected since, as is well known, extrapolation is not reliable over a large range of the independent variable. 
In fig.~\ref{fig-ML1} we show $A_{TL}$ at 10th and 12th order using both the extrapolated and machine learned coefficients. One sees that the extrapolated 10th order result lies within the error bar of the learned result, but the 12th does not despite the fact that, as expected, the error bar of the machine learned result increases at higher orders. 

\begin{figure}[H]
\begin{center}
\includegraphics[width=15cm]{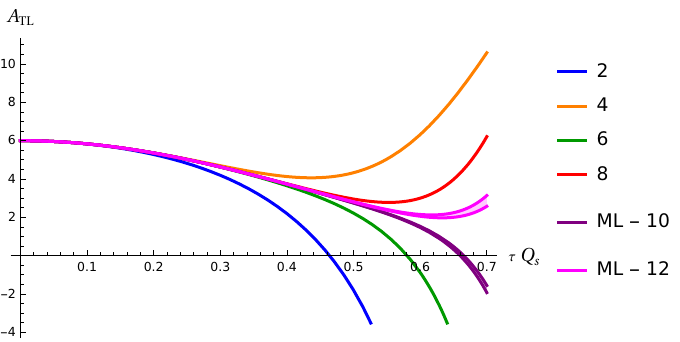}
\end{center}
\caption{$A_{TL}$ versus $\tau Q_s$ up to 12th order. 
\label{fig-ML2}}
\end{figure}

In fig.~\ref{fig-ML2} we show the graph of $A_{TL}$ to 12th order including the two machine learned coefficients. The graph shows the expected convergence properties of the expansion. The largest time at which the 12th order result can be trusted is approximately $\tau\approx 0.65$~fm.

To further test our method we have tried to learn the ratios in eq.~(\ref{magic}) instead of using them. 
To do this we repeat the process we used to learn the numerator, but this time we learn the denominator. 
The coefficients of the polynomial to eighth order are
\bea
\{d_0,d_2,d_4,d_6,d_8\} = \{15.9773,  -29.6759,  42.6822, - 49.2686,  50.4716\}\,.
\label{us8den}
\eea
We use $d_{10}=-47\pm 3$ and $d_{12}=50\pm 4$  for the base numbers of the unknown coefficients $d_{10}$ and $d_{12}$ and multiply by random numbers from a normal distribution as before.
%
We remove any results for which the absolute value of any coefficient is greater than $10^{7}$. 
To evaluate the success of the calculation we find the ratios $n_i/d_i$ and compare with eq.~(\ref{magic}). 
We calculate both the average and median of the learned ratios. The results are shown in table~\ref{make-ratio}. 
The median values agree with the theoretical results to fairly high accuracy up to 12th order. As expected, the average values are reliable only for small values of $i$. 
\begin{table}[H]
\begin{center}
\begin{tabular}{|c|c|c|c|c|}
\hline
i & $n_i/d_i$ & ~~ median ~~  & average \\
\hline\hline
0 & 6 & 6. & 6. \\
2 & 15 & 15. & 15.000$\pm$ 0.035 \\
4 & 24 & 24.000000$\pm$ 0.000017 & 24.0$\pm$ 0.4 \\
6 & 33 & 33.00000$\pm$ 0.00007 & 35.$\pm$ 6. \\
8 & 42 & 42.00000$\pm$ 0.00015 & 33.$\pm$ 16. \\
10& 51 & 50.9$\pm$ 0.8 & 41.$\pm$ 18. \\
12& 60 & 59.$\pm$ 6. & -270.$\pm$ 624. \\
 \hline
 \end{tabular}
 \end{center}
 \caption{The values of $n_i/d_i$ and the median and average of the learned ratios. 
 \label{make-ratio}}
\end{table}

\section{Discussion and Conclusions}

Proper time expansions can be used to obtain analytic results from a CGC effective theory.  
A  disadvantage of this method is that results are valid only at very early times. Since calculations are not feasible beyond eighth order in the expansion, the usefulness of these calculations is somewhat limited. We have studied three different methods to extend the upper bound at which different quantities can be calculated. 

In sec.~\ref{sec-ming} we have studied the method of Li and Kapusta \cite{Li:2016eqr}. The approximation is based on the existence two different ultraviolet scales, one that sets the upper bound for the validity of the classical approximation and another, smaller but still hard, which is the saturation scale. The calculation of the EMT is vastly simplified by including only terms that are leading order, and next-to-leading order, in the ratio of these scales. 
We have shown however that the LK approximation is not suited to calculate quantities that depend on nuclear structure because the approximation suppresses the relevant terms in the EMT. 
In sec.~\ref{sec-pade} we have used Pad\'e approximants to extend some full eighth order results to longer times. We have estimated the reliability of the results by using several different approximants and  different methods to select the data used to calculate the coefficients. 
In sec.~\ref{sec-ML} we have explored the use of machine learning techniques.
We have developed a method to use the package \textbf{PySR }\cite{Cranmer}. We have demonstrated that our method can learn the known eighth order coefficients of the transverse-longitudinal pressure anisotropy assuming knowledge of only the lower order ones. We have applied the same method to learn the 10th and 12th order coefficients and argued that the errors in these results is less than about 5\%. 

\begin{figure}[h]
\begin{center}
\includegraphics[width=11cm]{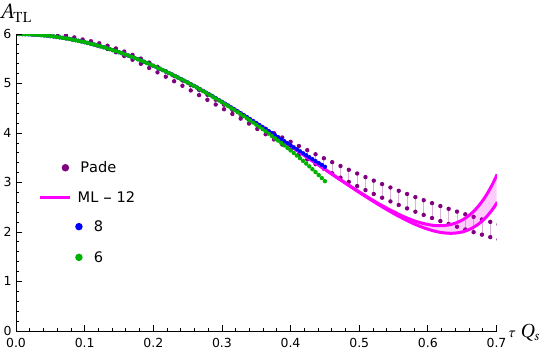}
\end{center}
\caption{A comparison of the results for $A_{TL}$ from the fourth order Pad\'e approximant and the 12th order maching learning calculation. 
\label{fig-compare}}
\end{figure}

We have calculated the transverse-longitudinal pressure anisotropy (see eq.~(\ref{Atl-def})) using all three of the methods studied in this paper. This quantity is of particular interest for several reasons. It is physically important because it shows the extent to which the system has approached equilibrium. 
It depends on ratios of elements of EMT and therefore is expected to be insensitive to the proportionality factor in the equation that relates the MV scale to the saturation scale, which cannot be determined in the CGC approach that we use. It is especially difficult to estimate $A_{TL}$ using extrapolation methods because the proper time expanded results 
diverge in opposite directions at each successive order in the approximation. 
The results for the maximum time at which $A_{TL}$ can be calculated, using the full EMT and the three approximate methods that we have studied in this paper, are summarized in table \ref{results-fin}. The maximum time that can be reached is increased by a factor of approximately 1.5.
\begin{table}[H]
\begin{center}
\begin{tabular}{|c|c|c|c|c|}
\hline
 $\tau_{\rm max}$ [fm] & ~~~ full ~~~ & ~~~ LK ~~~ & ~~~ Pad\'e ~~~  & ~~~ ML ~~~ \\
\hline
& 0.05 & 0.08 & 0.08 & 0.065 \\
 \hline
 \end{tabular}
 \end{center}
\caption{The approximate maximum time at which the transverse-longitudinal pressure anisotropy can be calculated using a proper time expansion of the full eighth order EMT and the three approximation methods we have studied.  \label{results-fin}}
\end{table}
We cannot directly compare the results of all three calculations because the LK approach depends on the assumption that there are two relevant ultra-violet scales and the other two calculations use only one ultra-violet scale. In fig.~\ref{fig-compare} we show the transverse-longitudinal pressure anisotropy from the Pad\'e approximant and the machine learning calculation. The plot shows the purple curve from fig.~\ref{plot-padeA} and the magenta curve from fig.~\ref{fig-ML2}, together with the analytic results at sixth and eighth order. The graph shows that the machine learned result appears to follow the data more closely initially, but the Pad\'e result maintains a more physically realistic form at longer times.  

We expect that Pad\'e approximants could be used to extend the radius of convergence of the proper time expansion by roughly 35\%, largely independently of the quantity being calculated. The LK approximation is expected to be about as good as a Pad\'e, but only assuming a large separation between the two ultraviolet scales and only for quantities for which nuclear structure is not important. The machine learning techniques we have investigated are not quite as efficient, but these methods are in their infancy and there is no doubt that the possibilities to improve their application to problems of this type are enormous.

\end{document}